\documentclass[twoside,twocolumn,9pt]{article}
\usepackage{extsizes}
\usepackage[super,sort&compress,comma]{natbib} 
\usepackage[version=3]{mhchem}
\usepackage[left=1.5cm, right=1.5cm, top=1.785cm, bottom=2.0cm]{geometry}
\usepackage{balance}
\usepackage{mathptmx}
\usepackage{sectsty}
\usepackage{graphicx} 
\usepackage{lastpage}
\usepackage[format=plain,justification=justified,singlelinecheck=false,font={stretch=1.125,small,sf},labelfont=bf,labelsep=space]{caption}
\usepackage{float}
\usepackage{fancyhdr}
\usepackage{fnpos}
\usepackage[english]{babel}
\addto{\captionsenglish}{%
  
}
\usepackage{array}
\usepackage{droidsans}
\usepackage{charter}
\usepackage[T1]{fontenc}
\usepackage[usenames,dvipsnames]{xcolor}
\usepackage{setspace}
\usepackage[compact]{titlesec}

\definecolor{cream}{RGB}{222,217,201}

\usepackage{siunitx}
\usepackage{upgreek}
\usepackage{amsmath}
\usepackage{amssymb}
\usepackage{soul}
\usepackage{xr}
\newcommand{\vdel}{\ensuremath{\mathbf{v}_{\!\Delta}}}
\newcommand{\vp}{\ensuremath{\mathbf{v}_{\mathrm{p}}}}
\newcommand{\vf}{\ensuremath{\mathbf{v}_{\mathrm{f}}}}

\DeclareSIUnit\molar{M}

\begin{document}

\pagestyle{fancy}
\thispagestyle{plain}
\fancypagestyle{plain}{
\renewcommand{\headrulewidth}{0pt}
}

\makeFNbottom
\makeatletter
\renewcommand\LARGE{\@setfontsize\LARGE{15pt}{17}}
\renewcommand\Large{\@setfontsize\Large{12pt}{14}}
\renewcommand\large{\@setfontsize\large{10pt}{12}}
\renewcommand\footnotesize{\@setfontsize\footnotesize{7pt}{10}}
\makeatother

\renewcommand{\thefootnote}{\fnsymbol{footnote}}
\renewcommand\footnoterule{\vspace*{1pt}%
\color{cream}\hrule width 3.5in height 0.4pt \color{black}\vspace*{5pt}} 
\setcounter{secnumdepth}{5}

\makeatletter 
\renewcommand\@biblabel[1]{#1}            
\renewcommand\@makefntext[1]%
{\noindent\makebox[0pt][r]{\@thefnmark\,}#1}
\makeatother 
\renewcommand{\figurename}{\small{Fig.}~}
\sectionfont{\sffamily\Large}
\subsectionfont{\normalsize}
\subsubsectionfont{\bf}
\setstretch{1.125} 
\setlength{\skip\footins}{0.8cm}
\setlength{\footnotesep}{0.25cm}
\setlength{\jot}{10pt}
\titlespacing*{\section}{0pt}{4pt}{4pt}
\titlespacing*{\subsection}{0pt}{15pt}{1pt}

\fancyfoot{}
\fancyfoot[RO]{\footnotesize{\sffamily{1--\pageref{LastPage} ~\textbar  \hspace{2pt}\thepage}}}
\fancyfoot[LE]{\footnotesize{\sffamily{\thepage~\textbar\hspace{3.45cm} 1--\pageref{LastPage}}}}
\fancyhead{}
\renewcommand{\headrulewidth}{0pt} 
\renewcommand{\footrulewidth}{0pt}
\setlength{\arrayrulewidth}{1pt}
\setlength{\columnsep}{6.5mm}
\setlength\bibsep{1pt}

\makeatletter 
\newlength{\figrulesep} 
\setlength{\figrulesep}{0.5\textfloatsep} 

\newcommand{\topfigrule}{\vspace*{-1pt}%
\noindent{\color{cream}\rule[-\figrulesep]{\columnwidth}{1.5pt}} }

\newcommand{\botfigrule}{\vspace*{-2pt}%
\noindent{\color{cream}\rule[\figrulesep]{\columnwidth}{1.5pt}} }

\newcommand{\dblfigrule}{\vspace*{-1pt}%
\noindent{\color{cream}\rule[-\figrulesep]{\textwidth}{1.5pt}} }

\makeatother

\newcommand{\pacorr}[1]{{#1}}
\newcommand{\pa}[1]{{#1}}
\newcommand{\JH}[1]{\textcolor{magenta}{JENS: #1}}
\newcommand{\david}[1]{\textcolor{ForestGreen}{DAVID: #1}}



\twocolumn[
  \begin{@twocolumnfalse}
\vspace{1em}
\sffamily
\begin{tabular}{m{4.5cm} p{13.5cm} }

& \noindent\LARGE{\textbf{Hydrodynamic simulations of sedimenting dilute particle suspensions under repulsive DLVO interactions$^\dag$}} \\
\vspace{0.3cm} & \vspace{0.3cm} \\

 & \noindent\large{David Jung,\textit{$^{ab}$} Maximilian Johannes Uttinger,\textit{$^{cd}$}, Paolo Malgaretti,\textit{$^{a}$} Wolfgang Peukert,\textit{$^{cd}$} Johannes Walter\textit{$^{cd}$} and Jens Harting\textit{$^{aeb\ast}$}} \\
& \noindent\normalsize{
We present guidelines to estimate the effect of electrostatic repulsion in
	sedimenting dilute particle suspensions. Our results are based on
	combined Langevin dynamics and lattice Boltzmann simulations for a
	range of particle radii, Debye lengths and particle concentrations.
	They show a simple relationship between the slope $K$ of the
	concentration-dependent sedimentation velocity 
	and the range $\chi$
	of the electrostatic repulsion normalized by the average
	particle-particle distance. When $\chi \to 0$, the particles are too
	far away from each other to interact electrostatically and $K=6.55$ as
	predicted by the theory of Batchelor. As $\chi$ increases, $K$
	likewise increases as if the particle radius increased in proportion to $\chi$ up to a maximum around $\chi=0.4$. Over the range $\chi=0.4-1$, $K$ relaxes exponentially to a concentration-dependent constant consistent with known results for ordered particle distributions.
	Meanwhile the radial distribution function transitions from a disordered gas-like
	to a liquid-like form.
        Power law fits to the concentration-dependent sedimentation velocity similarly yield a simple master curve for the exponent as a function of $\chi$, with a step-like transition from 1 to 1/3 centered around $\chi = 0.6$.
} \\

\end{tabular}

 \end{@twocolumnfalse} \vspace{0.6cm}

  ]

\renewcommand*\rmdefault{bch}\normalfont\upshape
\rmfamily
\section*{}
\vspace{-1cm}


\footnotetext{\textit{$^{a}$~Helmholtz Institute Erlangen-N{\"u}rnberg for Renewable Energy, Forschungszentrum J{\"u}lich, F{\"u}rther Stra{\ss}e 248, 90429 N{\"u}rnberg, Germany.}}
\footnotetext{\textit{$^{b}$~Department of Physics, Friedrich-Alexander-Universit{\"a}t Erlangen-N{\"u}rnberg, F{\"u}rther Stra{\ss}e 248, 90429 N{\"u}rnberg, Germany.}}
\footnotetext{\textit{$^{c}$~Institute of Particle Technology (LFG), Friedrich-Alexander-Universit{\"a}t Erlangen-N{\"u}rnberg (FAU), Cauerstra{\ss}e 4, 91058 Erlangen, Germany.}}
\footnotetext{\textit{$^{d}$~Interdisciplinary Center for Functional Particle Systems (FPS), Friedrich-Alexander-Universit{\"a}t Erlangen-N{\"u}rnberg, Haberstra{\ss}e 9a, 91058 Erlangen, Germany.}}
\footnotetext{\textit{$^{e}$~Department of Chemical and Biological Engineering, Friedrich-Alexander-Universit{\"a}t Erlangen-N{\"u}rnberg, F{\"u}rther Stra{\ss}e 248, 90429 N{\"u}rnberg, Germany. E-mail: j.harting@fz-juelich.de}}

\footnotetext{\dag~Electronic Supplementary Information (ESI) includes several auxiliary plots and short derivations. See DOI: 00.0000/00000000.}



\section{Introduction}
The physics of sedimenting particles have proven to be surprisingly difficult
to model despite many attempts over a large fraction of the 20th century.  While a
single particle slowly sedimenting in a sufficiently large container can be
easily described by Stokes' law, the long-ranged nature of hydrodynamic
interactions renders the dependence of the sedimentation speed on the particle
concentration complicated to derive even in the dilute limit.

For the purpose of brevity we refer to the case of uncharged particles
interacting only via hydrodynamic and hard sphere interactions as the case of
non-interacting particles throughout this paper.
The theory of non-interacting particles reached a major
breakthrough when in 1972 Batchelor\cite{batchelor_sedimentation_1972} derived
the sedimentation velocity $v$ at small particle volume fractions
$\phi$ relative to the velocity $v_0$ at infinite dilution as
\begin{equation}
  \label{eq:basic_batchelor}
\frac{v}{v_0}=1- K \phi,
\end{equation}
with $K=6.55$.
Similarly, the sedimentation velocity is sometimes written as
\begin{equation}
  \label{eq:inverse_sedimentation}
  \frac{v}{v_0} = \frac{1}{1 + K \phi},
\end{equation}
which is identical to Eq.~\eqref{eq:basic_batchelor} in the limit of small $\phi$.
The sedimentation velocity remains positive for all $\phi$ following Eq.~\eqref{eq:inverse_sedimentation}, unlike Eq.~\eqref{eq:basic_batchelor}, though neither equation is accurate anywhere near the concentration $\phi \approx 15\%$ where Eq.~\eqref{eq:basic_batchelor} goes to zero.
Beyond the dilute limit, the Rotne-Prager far-field approximation of hydrodynamic interactions was shown by Brady and Durlofsky in 1988 to be accurate for non-interacting spheres even up to a volume fraction of 50\%.\cite{brady_sedimentation_1988} \pacorr{Alternatively, via a Virial expansion, Cichoki et al. attempted to take into account three-particle contributions.~\cite{Cichocki2002}}
Experimentally, good agreement with Eq.~\eqref{eq:basic_batchelor} has been shown to require a P\'eclet number $\mathrm{Pe} < 1$ due to its underlying assumption of a perfectly homogenous radial distribution function (RDF).\cite{benes_sedimentation_2007}

Especially for particles of nanometer scale
neglecting any non-hydrodynamic
interparticle interactions is a strict limitation though. Indeed, depending on the pH
value, most
types of colloidal particles tend to accumulate considerable surface charges
when dissolved in water\cite{shaw_introduction_1980,horn_surface_1978,ohsawa_zeta_1986}. 
This
leads to strong electrostatic interactions which typically decay over a Debye
length of the order of $\SI{10}{\nm}$. The Debye length in water can in
principle reach hundreds of nanometers, though this requires high degrees of
purity that are in practice difficult to achieve.\\
For this
reason a majority of studies on the sedimentation of interacting particles focus
on attractive potentials.\cite{fiore_hindered_2018,sun_investigating_2018} In
organic solvents such as ethanol, however, Debye lengths of around
$\SI{800}{\nm}$ have been reached in
experiments.\cite{thies-weesie_nonanalytical_1995}

For particle suspensions with strong electrostatic interactions and weak
screening (i.e.~a 
Debye length $\lambda_\mathrm{D}$ large enough to be comparable to the
average particle-particle distance) a strongly nonlinear decrease of the
sedimentation velocity with concentration has been both predicted theoretically
and observed
experimentally\cite{thies-weesie_nonanalytical_1995,gilleland_new_2011} even in
the dilute limit where $\phi<1\%$.

Early studies of electrostatic effects in particle sedimentation include the work of
Booth\cite{booth_sedimentation_1954} in 1954.
They developed the dipole moment
of sedimenting charged particles as a power series in terms of the particle
charge or zeta potential and managed to calculate the first two coefficients in
the series.
The theory is thus appropriate for sufficiently low surface charges/zeta
potentials, although this limitation was removed in a numerical extension of
Booth's work by Stigter in 1980.\cite{stigter_sedimentation_1980} 
Both Booth's and Stigter's theories completely neglect hydrodynamic
interactions between the particles and do not take changes in the RDF of the
suspension into account.

A number of
studies\cite{levine_prediction_1976,ohshima_sedimentation_1998,ding_sedimentation_2001,lee_sedimentation_2002}
of sedimentation under both electrostatic and hydrodynamic particle-particle
interactions have been performed using  
methods based on geometric cells to obtain the
hydrodynamic component, either with the free-surface boundary condition by
Happel\cite{happel_viscous_1958} or the zero vorticity condition by
Kuwabara.\cite{kuwabara_forces_1959}
While experimental results confirm the cell models
as adequate to calculate the
sedimentation potential,\cite{marlow_sedimentation_1985} both the method by
Happel and that of Kuwabara fail to correctly reproduce the sedimentation
behavior of non-interacting particles in the dilute limit found by Batchelor
about 14 years after the introduction of the
method.\cite{happel_viscous_1958,kuwabara_forces_1959,batchelor_sedimentation_1972}
Furthermore, the methods based on geometric cells cannot take into account changes in the RDF of
the sedimenting suspension induced by the electrostatic interactions and they
assume an electrically neutral unit cell, which may be a too rough
simplification if Debye layers overlap
strongly.\cite{nagele_electrokinetic_2013}

Another promising approach in modeling charged particle sedimentation
numerically was taken by Watzlawek and
N{\"agele},\cite{watzlawek_sedimentation_1999} though their approach is limited
by the fact that it can only take into account pair-wise hydrodynamic 
interactions.
Neglecting many-body hydrodynamic interactions was shown by Brady and Durlofsky\cite{brady_sedimentation_1988} to lead to a significant error in the sedimentation rate at volume fractions as low as 5\%,
 though the result could be improved considerably by additionally neglecting stresslet contributions as per the Rotne-Prager approximation.
  Approximate many-body hydrodynamic interactions can be taken into account
  using the Stokesian dynamics method\cite{brady_stokesian_1988}
  and 
  advancements in recent years have improved its performance up to
  a linear scaling with the number of particles.\cite{fiore_fast_2019}
  Nonetheless, the
  handling of hydrodynamic interactions remains fundamentally approximate in Stokesian dynamics due to
  a truncated expansion of the mobility matrix.
  Furthermore, the method is limited in terms of its extensibility to non-zero Reynolds numbers and polydisperse or non-spherical particles.
  Parallelized Stokesian dynamics implementations scale efficiently to up to a few hundred CPUs\cite{bulow_scalable_2016} and have been used to study the sedimentation of aggregates of thousands of polydisperse particles.\cite{bulow_settling_2015}

Banchio et al. and
  Gapinski et
  al.,\cite{banchio_many-body_2006,gapinski_collective_2007,banchio_short-time_2008}
  have previously employed the Stokesian dynamics method to numerically study suspensions
  under repulsive interactions.
  They obtained the structure factor of the suspension and the so-called hydrodynamic function
  $H(q)$ for selected values of salt and particle concentrations. Though
  their results are focused more on modelling diffusivity, the hydrodynamic
  function
  contains the relative sedimentation speed of the suspension under
  a spatially constant force for $q=0$. 
  Comparison of experiments
  with the hydrodynamic function for a given concentration and as a function of
  $q$ requires measuring the static structure factor, e.g.
  via X-ray scattering, as well as the collective diffusion function, e.g. via
  dynamic light scattering. Our approach of quantifying the functional shape and
  the mean slope of the sedimentation velocity as a function of concentration
  for a broad range of salt concentrations and different particle concentration
  ranges should lend itself to a more straightforward comparison to centrifugal
  sedimentation experiments.
In fact we have recently applied an early version of our method described in this work in an experimental context.\cite{uttinger_probing_2021}

As an alternative to Stokesian dynamics one can model the sedimentation of
  particles in a fluid by coupling the discrete element method for
  the dynamics of the suspended particles to a Stokes or Navier-Stokes level hydrodynamics solver.
  Many different methods
  have been used for the latter, such as directly solving the
  Navier-Stokes equation using the finite element
  method,\cite{gan_direct_2003} smoothed particle
  hydrodynamics,\cite{robinson_grain_2013} or stochastic rotation
  dynamics.\cite{perez_comparative_2015,schafer_agglomeration_2010} In this work we employ the
  lattice Boltzmann method (LBM). It has been shown to be a viable
  tool to capture the full hydrodynamic interactions of large numbers
  of non-interacting sedimenting particles by Nguyen and Ladd in
  2005,\cite{nguyen_sedimentation_2005} though there is similar work
  by Ladd with smaller particle numbers dating back to
  1994.\cite{ladd_numerical_1994} Later on the method has similarly
  been used to model particles with attractive interaction
  potentials.\cite{derksen_simulations_2014} Several different
  algorithms for coupling particles to the LBM fluid exist, the method
  is numerically efficient and is not limited to low Reynolds number
  flows.\cite{gao_lattice_2013} For low Reynolds number flows the LBM
  has been found to give results consistent with the Stokesian
  dynamics method.\cite{binder_simulation_2006,schlauch_comparison_2013}

In this paper we numerically study the impact of electrostatic interactions
modeled by DLVO theory on sedimenting suspensions under varied particle size, concentration and Debye length.
By simulating the interactions of
a large number of particles and the resulting changes in the RDF explicitely and
by including full hydrodynamic interactions using the LBM we improve upon
previous studies and contribute to a clearer picture of how electrostatic
interactions influence particle sedimentation.\\

\section{Model and methodology}
\label{sec:methodology}
Each sedimentation simulation for a given set of concentration, particle size,
and Debye length parameters consists of two major steps. First, a set of
particle positions representative of an equilibrated bulk suspension of charged
particles interacting via DLVO potentials is generated from a Langevin dynamics
simulation.  Second, the hydrodynamic interactions and the resulting
sedimentation velocity under added constant acceleration (representing
gravitational or centrifugal forces) are calculated for the particle positions
obtained previously using the LBM. The final result is the particle velocity in
the direction of the constant acceleration averaged over all particles.

While in the first step both particle positions and particle velocities evolve
in time, only the velocities are updated in the last step while the positions
remain fixed. In this way we neglect changes in the RDF induced by
hydrodynamic interactions and greatly reduce the convergence time and numerical
cost of our hydrodynamic simulations. We consider this simplification to be
justified in the limit of small P\'eclet numbers, where particle advection
plays a small role compared to diffusion and drift fluxes induced by strong
DLVO interactions.  As we keep the particle positions fixed, neither advection
nor diffusion occur in our hydrodynamic simulations so that the P\'eclet number
is not obviously defined.  However, by keeping the Reynolds number small
($\mathrm{Re} \lesssim \num{5e-6}$) we can consider the fluid flow velocity
around the particles to be arbitrarily small.  It follows that the P\'eclet
number $\mathrm{Pe} \propto \mathrm{Re} / D$ calculated using the diffusivity
$D$ of the particles in the preceding Langevin simulation is likewise
vanishingly small.

Ongoing research into the possible causes of an observed slow decay of sedimentation velocity fluctuations has led to the widespread assumption that subtle changes in the RDF may be taking place in sedimenting suspensions over long time spans up to several hours, even in the limit of small $\mathrm{Pe}$.\cite{koch_screening_1991,moller_velocity_2017}
Reproducing this experimentally observed decay of velocity fluctuations accurately would require significantly longer simulation times\cite{ladd_effects_2002} and the presence of confinement\cite{brenner_screening_1999} with a geometry matching the experimental system.\cite{tee_nonuniversal_2002}
To our knowledge, however, no corresponding long term evolution of the mean sedimentation velocity has been observed so far in monodisperse suspensions.

\subsection{Generating particle positions}
\label{sec:generating_particle}
In the first step, we initialize about \num{10000} spherical particles with
random positions $\mathbf{r}_i$ without overlap in a 3D rectangular system with
periodic boundary conditions. The particle positions are evolved in time $t$ in
each spatial dimension $\iota$ according to the Langevin equation
\begin{equation}
  \label{eq:langevin}
  m\frac{\partial^2 r^\iota_i}{\partial t^2} = -\gamma \frac{\partial r^\iota_i}{\partial t} + \sum_{j\neq i} {F}_i^\iota(\mathbf{r}_i,\mathbf{r}_j)+\eta_i^\iota (t).
\end{equation}
The particle mass $m$ is set to reproduce a particle density of
\SI{1800}{\kg\per\m\cubed}, which is a realistic value for e.g.
SiO$_2$ nanoparticles. Pairwise particle interaction forces
$\mathbf{F}_i(\mathbf{r}_i,\mathbf{r}_j)$ account for DLVO and hard sphere interactions
with all surrounding particles up to a cutoff radius carefully
selected depending on the range of the DLVO interactions. Stokes' law
provides the translational friction coefficient $\gamma=6\pi\mu R$
based on the dynamic viscosity $\mu$. The randomized force $\eta$
represents thermal fluctuations and fulfills the fluctuation
dissipation theorem in each spatial dimension, which is given by
\begin{equation}
  \label{eq:fluctuationdissipation}
  \langle \eta^\iota(t)\eta^\iota(t') \rangle = 2\mathrm{\mathrm{k_BT}} \gamma \delta(t-t').
\end{equation}
The Langevin equation is discretized in time and the particle positions are
updated according to the leapfrog algorithm. Convergence of the Langevin
dynamics simulations is determined based on the time evolution of the total
DLVO interaction energy in the system.  When the drift in the energy over the
last \num{5000} time steps is smaller than the standard deviation of the energy
due to thermal fluctuations, the simulation is stopped.  The final particle
positions are then transferred to a lattice Boltzmann (LB) simulation to
determine the hydrodynamic interaction of the particles.  For simulations with
no DLVO interactions, the Langevin dynamics simulations are skipped and random
particle positions are used in the LB simulation.

\subsection{Particle-fluid coupling}
\label{sec:particle_fluid_coupling}
LB simulations are performed using our in-house code {\it
LB3D}.\cite{schmieschek_lb3d_2017,JTH10,HCVC05} In our LB implementation,
fluid properties are calculated on a regular cubic lattice in three dimensions.
On each lattice site 19 scalar populations $f_i$ are defined and the lattice
constant is referred to as $\Delta x$. Each population is proportional to the
fraction of fluid flowing in the velocity direction $\mathbf{c}_i$ toward a neighboring lattice
site or remaining at rest ($\mathbf{c}_{19}=0$) during that time step. In each time
step $\Delta t$ they are shifted to a neighboring lattice site according to
\begin{equation}
  \label{eq:lbm_streaming}
  f_i(\mathbf{r}, t) \to f_i(\mathbf{r}+\mathbf{c}_i{\Delta t}, t+\Delta t),
\end{equation}
and then relaxed toward an equilibrium distribution $f_i^{\mathrm{eq}}$ in the
collision step
\begin{equation}
  \label{eq:lbm_collision}
  f_i \to f_i - \frac{1}{\tau_{\mathrm{r}}}(f_i - f_i^{\mathrm{eq}}) + S_i.
\end{equation}
This approach, using a single (dimensionless) relaxation time $\tau_{\mathrm{r}}$, is known as the BGK
scheme, after Bhatnagar, Gross and Krook.\cite{bhatnagar_model_1954}.
$f_i^{\mathrm{eq}}$ is a truncated
Maxwell-Boltzmann distribution for the discretized set of possible velocities
along the velocity directions $\mathbf{c}_i$.\cite{BenSucSau92,kruger_lattice_2017} The source term $S_i$ stems from the action of body forces. A number of different
schemes to calculate $S_i$ have been shown to produce physically accurate
results. For this work we choose a scheme by
Kupershtokh.\cite{kupershtokh2004new,kupershtokh_equations_2009} Letting
$\Delta m$ be the unit mass, the fluid mass density $\rho_\mathrm{f}$ and
velocity $\vf$ at a given lattice site are calculated as
 \begin{align}
   \label{eq:lbm_moments-paolo}
   \rho_\mathrm{f} &= \Delta m \sum_{i=1}^{19} f_i, \\
 	\vf &= \frac{\Delta m}{\rho_\mathrm{f} }\sum_{i=1}^{19} f_i \mathbf{c}_i
 \end{align}
and the dynamic viscosity is given as
\begin{equation}
  \label{eq:lbm_viscosity}
  \mu = \frac{\rho_\mathrm{f}}{6} (2\tau_{\mathrm{r}} - 1) \frac{\Delta x^2}{\Delta t}.
\end{equation}
\pa{We performed all our simulations with $\tau_{\mathrm{r}} = 1$ for reasons of numerical simplicity. Changing the relaxation time and thus the viscosity only changes the overall time
scale of the system.}
Particles are coupled to the interpolated fluid velocity via a linear friction force
in an approach
based on work by
Ahlrichs and D{\"u}nweg.\cite{ahlrichs_lattice-boltzmann_1998}  According to
Stokes' law, the friction force experienced by a single particle of velocity
$\vp$ inserted into a fluid flowing with velocity $\vf$ is
\begin{equation}
  \label{eq:stokes_force}
\mathbf{F}_\mathrm{s}=-\gamma \big(\vp-\vf\big) = -\gamma \vdel.
\end{equation}
The same friction force, with the opposite sign, also acts on the fluid
following Newton's third law.
The effect of each particle on the fluid flow is limited to the effect of this point-like friction force, yielding the Stokeslet approximation of its full flow field.
As a result, particle rotation is not included in the model.\\
Applying Eq.~\eqref{eq:stokes_force}
in the numerical model using $\gamma=6\pi \mu R$ and setting $\vf$ equal to the
fluid velocity from the LBM interpolated to the particle position results in
steady-state velocities $v_{\mathrm{p}}=|\vp|$ that are higher than the expected
result from Stokes' theory. This is because $\vf$ in Eq.~\eqref{eq:stokes_force} represents
the fluid velocity without the Stokeslet contribution from the considered particle according to Stokes' law.
\pa{We chose a particle radius equal to the LB grid spacing in order to have a relatively large radius while ensuring that the particle geometry remains comfortably within the extent of its stencil surrounding it. Furthermore we verified that the friction force densities remain smaller by more than an order of magnitude at all times compared to values deemed problematic in the lattice Boltzmann method.}
Fortunately, the contribution of the particle to its surrounding flow field can be
easily subtracted by rescaling the friction coefficient, as shown by Ollila et
al.\cite{ollila_hydrodynamic_2013}
\begin{equation}
  \label{eq:friction_rescaling}
  \gamma \longrightarrow \left( \frac{1}{\gamma} - \frac{1}{\gamma_s} \right)^{-1}.
\end{equation}
The correction factor $\gamma_s$ can be determined
analytically in principle,\cite{nash_singular_2008} but it depends non-trivially on the stencil used to interpolate the fluid
velocity as well as other details of the numerical
implementation. 
Instead we choose the simpler approach of deriving $\gamma_s$ from fits to a series of
numerical measurements of the steady-state single particle velocity as a function
of the input friction coefficient.\cite{ollila_hydrodynamic_2013} Using a
cubic stencil with a side length of four lattice discretization lengths
$\Delta x$ and a weighting function derived by
Peskin,\cite{peskin_immersed_2002} we obtain
$\gamma_s \approx 4.91$ in simulation units. The corresponding fit is shown in Fig.~\ref{fig:si_frictionfit} in the ESI$^\dag$.
From here on, $\gamma$ always refers to the corrected friction coefficient
according to the substitution in Eq.~\eqref{eq:friction_rescaling}.

The sedimentation of particles with mass $m$ in our LB
simulations is triggered by a constant force $\mathbf{F}_\mathrm{g}$ representing
gravitational or centrifugal acceleration as well as the counteracting
buoyancy. The same force $\mathbf{F}_\mathrm{g}$ with opposite sign is distributed
homogenously among all fluid sites in the system. This ensures global
momentum conservation and mimics the backflow of displaced fluid occuring during sedimentation in a closed cell.

Assuming a constant $\vf$, the
particle velocity update by one time step due to the friction force alone can be written as
\begin{equation}
  \label{eq:particle_velocity}
  \vdel(t_{i+1}) = \vdel(t_{i}) + \Delta t \frac{\mathbf{F}_\mathrm{s}(t_i)}{m} = \vdel(t_{i}) \left( 1 - \Delta t \frac{\gamma}{m} \right).
\end{equation}
If $\Delta t \frac{\gamma}{m} < 1$, $\vp$ approaches $\vf$ via an exponential
decay. If $1 < \Delta t \frac{\gamma}{m} < 2$, $\vp$ oscillates
around $\vf$ due to discretization errors, but $|\vdel|$ still decays to zero
in time.  If, however, $\Delta t \frac{\gamma}{m} > 2$, then $|\vdel|$ diverges
to infinity in an oscillating manner.  The easiest way to avoid these
discretization effects would be to choose the time step such that $\Delta t <
m/\gamma$, or, at least, $\Delta t < 2m/\gamma$.  However, large values of
both $\gamma$ and $\Delta t$ are desirable when simulating a suspension at
low Reynolds number.
In order to avoid this issue, we analytically integrate the friction force $\mathbf{F}_{\mathrm{s}}$ under the assumption of a constant $\vf$ but continuously varying $\vp$ and $\mathbf{F}_\mathrm{s}$ over one time step, add the constant $\mathbf{F}_\mathrm{g}$, and calculate the average total force $\langle \mathbf{F}_\mathrm{T} \rangle_{\Delta t}$ as
\begin{equation}
  \label{eq:total_force}
  \begin{split}
  &\langle \mathbf{F}_\mathrm{T} \rangle_{\Delta t}(t_i) = \mathbf{F}_\mathrm{g}(t_i) - \frac{\gamma}{\Delta t} \int\displaylimits_{t_{i-\frac{1}{2}}}^{t_{i+\frac{1}{2}}} \vp(\tau)-\vf(t_i) \, \mathrm{d}\tau \\
    &=\bigg(1-\mathrm{e}^{-\frac{\gamma}{m}\Delta t}\bigg) \bigg(\frac{\mathbf{F}_\mathrm{g}(t_i)}{\gamma}-\vp(t_{i-\frac{1}{2}})+\vf(t_{i})\bigg) \frac{m}{\Delta t}.
  \end{split}
\end{equation}
Because we use the leapfrog algorithm to generate particle trajectories,
$\vp(t_{i-\frac{1}{2}})$ shifted by half a time step with respect to positions
and forces is readily available.  The fluid velocity in
Eq.~\eqref{eq:total_force} is $\vf(t_{i})$ instead of $\vf(t_{i-\frac{1}{2}})$
because we require the fluid velocity averaged over the time step from
$t_{i-\frac{1}{2}}$ to $t_{i+\frac{1}{2}}$.
In the overdamped limit, when $ m/\gamma \ll \Delta t$, Eq.~\eqref{eq:total_force} gives the same acceleration from $\mathbf{F}_{\mathrm{g}}$ as predicted by Brownian dynamics, plus advection by $\vf$. 

The averaged friction force acting on the fluid can be identified as $-(\langle
\mathbf{F}_\mathrm{T} \rangle_{\Delta t}-\mathbf{F}_\mathrm{g})$ and it is distributed to the
fluid sites surrounding the particle on the same stencil on which the
interpolation of $\vf$ takes place.

\begin{figure*}
  \centering
  \includegraphics{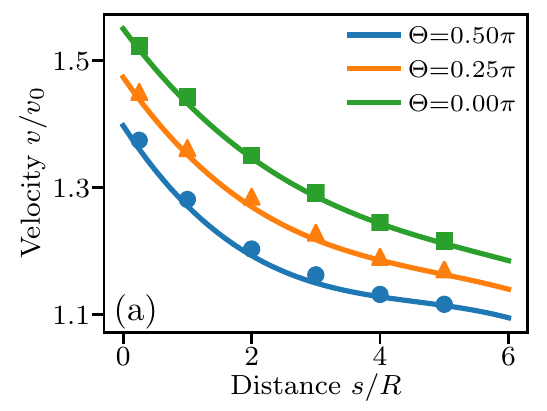}
  \includegraphics{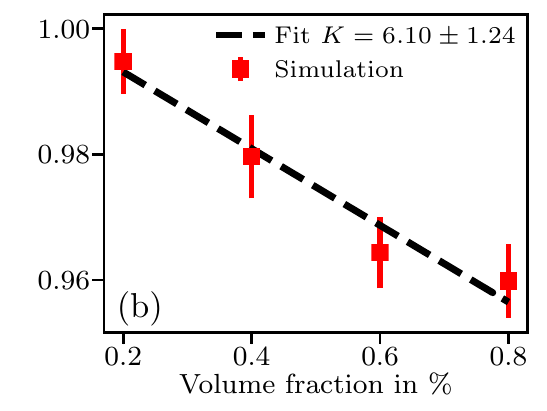}
  \includegraphics{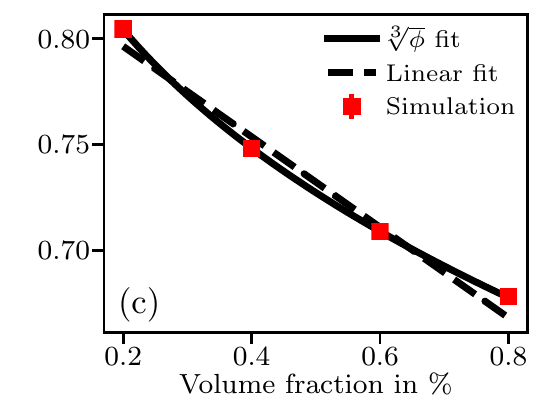}
  \caption{Sedimentation velocity in three different types of systems.
    (a) A pair of particles at fixed
	distance and with $\mathbf{F}_\mathrm{g}$ acting at angle $\Theta$
	to the connecting line between the particles. Full lines show the
	theory prediction following Eq.~\eqref{eq:pair_sedimentation}.
    (b) Suspension of non-interacting
	particles. Error bars stem from averaging over 6 simulations per
	concentration with different random particle placements. The dashed
	line shows the analytical solution by Batchelor.\cite{batchelor_sedimentation_1972}
    (c) Suspension under long-ranged repulsive DLVO interactions. The full line is a fit of the
	form $v/v_0=a-\varsigma\phi^{\frac{1}{3}}$ similar to
	Eq.~\eqref{eq:ordered_sedimentation}, giving $a=1.02$ and
	$\varsigma=1.71$. The dashed line is a linear fit yielding $K=21.3$.
        Error bars from averaging over 6 simulations are smaller than the symbols.
	}
  \label{fig:particle_benchmarks}
\end{figure*}
The LB simulations are considered converged when 
{the slope of the sedimentation
velocity over time relative to the velocity at infinite dilution and averaged
over all particles and the last 1000 time steps falls below a threshold value
of} \num{5e-8}.  This procedure usually requires between \num{5000} and
\num{20000} LB time steps.  We find that letting some simulations run up to
about thirty times longer changes the final sedimentation velocity by less than
0.01\%.

\subsection{DLVO interactions}
The total force acting on a particle in the Langevin simulations is calculated with the same averaging of the friction force introduced in Eq.~\eqref{eq:total_force} but setting the fluid velocity $\vf$ to zero and exchanging $\mathbf{F}_\mathrm{g}$ with a sum of DLVO and hard sphere pair potentials, i.e. $\mathbf{F}_\mathrm{g} \to \mathbf{F}_\mathrm{p}^i=\sum_j \mathbf{F}_\mathrm{DLVO}(|\mathbf{r}_i - \mathbf{r}_j|) + \mathbf{F}_\mathrm{hs}(|\mathbf{r}_i - \mathbf{r}_j|)$.
Each particle interacts only with particles within a numerical cutoff distance chosen according to the DLVO parameters of the simulation.
The DLVO interactions consist of an attractive contribution stemming from van der Waals interactions and a repulsive contribution stemming from Coulomb repulsion screened by counterions: $\mathbf{F}_\mathrm{DLVO}=\mathbf{F}_\mathrm{vdw}+\mathbf{F}_\mathrm{coul}$. The van der Waals force of two spheres of equal radius $R$ at a surface to surface distance $\hat{s} = s / R$ in multiples of $R$ is the derivative of the potential\cite{hunter_foundations_2001}
\begin{equation}
  \label{eq:vdw_potential}
    E_\mathrm{vdw} = -\frac{A_\mathrm{H}}{6}\bigg[ \frac{2}{\hat{s}^2+4\hat{s}} + \frac{2}{\hat{s}^2+4\hat{s}+4} + \mathrm{ln} \left( \frac{\hat{s}^2+4\hat{s}}{\hat{s}^2+4\hat{s}+4}\right)\bigg].
\end{equation}
We model van der Waals forces using an effective Hamaker constant of $A_\mathrm{H}=\SI{2e-20}{\joule}$.  This value is
similar to that measured by Fielden et al.\cite{fielden_oscillatory_2000} for
a silica particle interacting with a partially oxidized silicon wafer.
Valmacco et al.\cite{valmacco_dispersion_2016} measured substantially lower
values for pairs of silica particles in water, probably due to a high surface
roughness. A Hamaker constant of the order of $\SI{1e-20}{\J}$ is to be
expected for interactions between polystyrene particles in
water.\cite{watillon_interactions_1966}  As shown by example in Fig.~\ref{fig:si_reprange} of the ESI$^\dag$,
the strength of the repulsive component of the
DLVO interaction in the parameter space of large Debye lengths studied by us
renders the van der Waals interaction almost irrelevant for most of our simulations. We include van der Waals forces anyway for the sake of completeness.

The repulsive component consists of a Coulomb interaction between like-charged
spheres with an electrostatic potential $\zeta$ at the hydrodynamic slipping
plane, which is exponentially screened over a decay length $\lambda_\mathrm{D}$
by the presence of dissolved ions in a solvent of dielectric
permittivity $\varepsilon$\cite{hunter_foundations_2001}
\begin{equation}
  \label{eq:coul_potential}
  E_\mathrm{coul} = 4\pi R\varepsilon \zeta^2 \mathrm{e}^{-\frac{R}{\lambda_\mathrm{D}}\hat{s}}/(\hat{s}+2).
\end{equation}
A comparison of the resulting total DLVO potential
$E_\mathrm{DLVO}=E_\mathrm{vdw}+E_\mathrm{coul}$ with $E_\mathrm{coul}$ alone for
$R=\SI{300}{\nano\meter}$, $\zeta=\SI{50}{\milli\volt}$ and different values of
$\lambda_\mathrm{D}$ is shown in Fig.~\ref{fig:si_reprange} in the ESI$^\dag$.

The simplified pair-wise interactions of DLVO theory are computationally efficient and allow
us to reach large particle numbers with acceptable computational effort.
However, this approach neglects the deformation of the Debye layer in the
presence of a hydrodynamic flow.
While taking this deformation into account
could be achieved by coupling the solver for the fluid and particle dynamics to
a solver for the Nernst-Planck
equation,\cite{kuron_2016,rivas_2018} the influence of such ion advection effects
becomes negligible when the ions' P\'eclet number $\lambda_\mathrm{D} v / D_\mathrm{i}$ is small.\cite{khair_strong_2018}
As established in section~\ref{sec:methodology}, we are concerned in this work with systems of small particle P\'eclet number and Debye lengths comparable in size to the particle radius. The ions' P\'eclet number can be considered to be smaller still, owing to the smaller size and therewith larger diffusivity $D_\mathrm{i}$ of the ions as compared to the particles.
A fully resolved double layer would furthermore yield a reduction of the sedimentation velocity due to the restoring dipole force acting on the particle when it is accelerated by $\mathbf{F}_\mathrm{g}$ out of the center of its ionic atmosphere.\cite{stigter_sedimentation_1980,ohshima_sedimentation_1984,keller_2010}
Because this so-called primary charge effect is also present in the sedimentation of a single particle, we assume its effect on the relative sedimentation speed $v/v_0$ to be negligible.

In order to avoid strongly overlapping particles due to the divergence of $E_\mathrm{vdw}$ at contact when $\lambda_\mathrm{D}$ is small and thermal fluctuations allow particles to cross the potential barrier posed by $E_\mathrm{coul}$, a hard sphere repulsion term of the form
\begin{equation}
  \label{eq:hertz_potential}
  E_\mathrm{hs} = k (2 R - c)^{5/2}
\end{equation}
based on Hertzian contact theory\cite{landau_theory_1970} is applied to particles at center-to-center distances $c < 2 R$. The stiffness $k$ is chosen empirically based on the conditions that it needs to be sufficiently large to avoid significant particle overlap but small enough to not lead to excessive particle acceleration due to time discretization.

\subsection{Validation}
In order to check the accuracy of the particle-fluid coupling, we compare our
simulations with known results for the sedimentation behavior of
non-interacting particles.  First we compute the velocity of a pair of
neighboring particles under constant acceleration in Stokes flow. Two
particles with a radius equal to the length of discretization of the LB solver
are initialized in a fully periodic system.  As described in
section~\ref{sec:particle_fluid_coupling}, $\mathbf{F}_\mathrm{g}$ is applied to each
particle in the same direction and $-2\,\mathbf{F}_\mathrm{g}$ is spread homogenously
over all fluid lattice sites.
The component of the final sedimentation velocity in direction of
$\mathbf{F}_\mathrm{g}$ and relative to the velocity of a single particle can be
written as
\begin{equation}
  \label{eq:pair_sedimentation}
  \frac{v}{v_0} = \lambda_1 \cos^2{\Theta}+\lambda_2 (1 - \cos^2{\Theta}),
\end{equation}
where $\lambda_1$ and $\lambda_2$ as such are given in tabulated form as a
function of the interparticle distance by
Batchelor,\cite{batchelor_sedimentation_1972} albeit the original computations
were performed by Stimson and Jeffery\cite{stimson_motion_1926} for
$\lambda_1$, and Goldman et al.\cite{goldman_slow_1966} for $\lambda_2$.
Here, $\Theta$ is the angle between the connecting line of the particle centers
and the direction of $\mathbf{F}_{\mathrm{g}}$.  As shown in
Fig.~\ref{fig:particle_benchmarks}(a), very good agreement with
Eq.~\eqref{eq:pair_sedimentation} is obtained even when the
interparticle distance from center to center is less than 3 discretization
lengths.
This is remarkable, as the limitation of the fluid-particle coupling to the Stokeslet level means that hydrodynamic interactions are strictly accurate only in the far-field.

Next, we benchmark Eq.~\eqref{eq:basic_batchelor} for the sedimentation
velocity of non-interacting particles in bulk by simulating about
\num{10000} sedimenting particles in the same way as in the previous test.  The
corresponding results in Fig.~\ref{fig:particle_benchmarks}(b) also show good
agreement with Eq.~\eqref{eq:basic_batchelor}, with a measured $K=6.10\pm
1.24$.
Fig.~\ref{fig:particle_benchmarks}(c) shows results from an identical set of sedimentation simulations
with radius $R = \SI{100}{\nm}$ and a Debye length $\lambda_{\mathrm{D}} = \SI{950}{\nm}$.
The long-ranged repulsive interactions in these simulations lead to a functional form of $v(\phi) \propto \sqrt[3]{\phi}$, in agreement with theoretical expectations.\cite{thies-weesie_nonanalytical_1995}

\section{Results}
\begin{figure}
  \centering
  \includegraphics{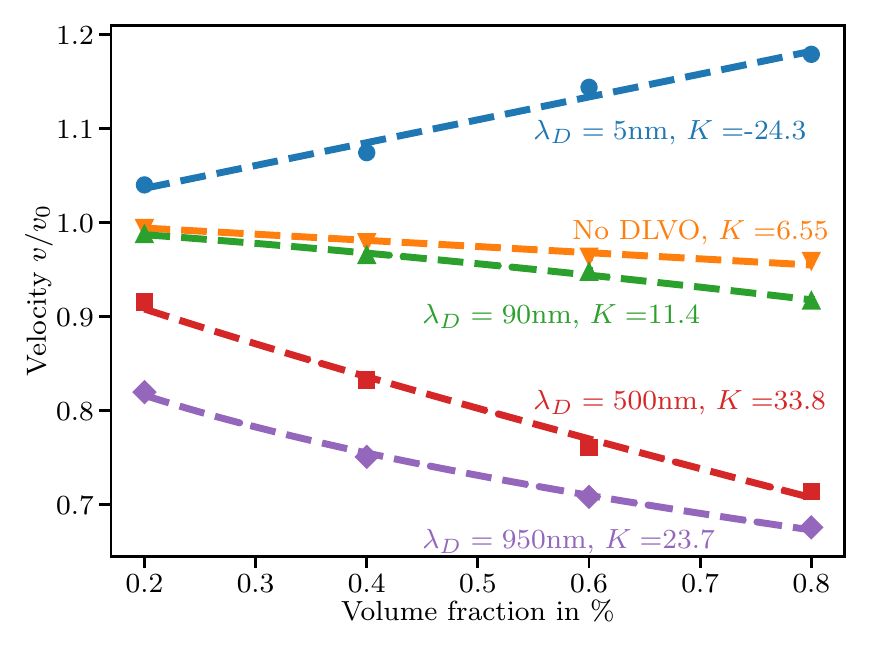}
  \caption{Sedimentation velocity in suspensions with particle radius $R=\SI{600}{\nm}$ and various Debye lengths.
    Aggregation causes the large positive slope at $\lambda_\mathrm{D}=\SI{5}{\nm}$.
    Dashed lines at $\lambda_\mathrm{D}=\SI{500}{\nm}$ and $\lambda_\mathrm{D}=\SI{950}{\nm}$ are nonlinear fits to Eq.~\eqref{eq:ordered_sedimentation} giving $\varsigma = 16.54,\,2.44$ and $\omega = 1.2,\,2.4$, respectively.}
  \label{fig:vel_phi}
\end{figure}
We perform simulations for a range of Debye lengths from $\SI{5}{\nano\meter}$
to 950 nm with particle radii $R$ set to 100, 200, 300, 450, and
$\SI{600}{\nm}$.  A Debye length of $\lambda_\mathrm{D}=\SI{5}{\nano\meter}$
corresponds to a monovalent salt concentration of about \SI{3.7}{\milli\molar}
in water at room temperature,  whereas $\SI{950}{\nm}$ is close to the Debye
length obtained in perfectly pure water solely by self-dissociation at a pH
value of 7. We keep the zeta potential fixed at $\zeta=\SI{50}{\milli\volt}$
regardless of the particle size.

For the smallest Debye length of \SI{5}{\nm} combined with the largest
particles of $R=\SI{600}{\nm}$, strong aggregation occurs, leading to negative
values of $K$, as shown in Fig.~\ref{fig:vel_phi}.  For smaller particles at
the same Debye length, we observe $K\approx 6.55$ and almost no aggregation.
To understand this, first note that the van der Waals potential in
Eq.~\eqref{eq:vdw_potential} does not depend on $R$ for a given $\hat{s}$.  The
repulsive potential in Eq.~\eqref{eq:coul_potential} on the other hand can be
shown in a simple mathematical exercise to always decrease when $R$ is
increased as long as $s>\lambda_\mathrm{D}$.  The proof can be found in
section~\ref{sec:si_radius_dependence} of the ESI$^\dag$.  Thus, the attractive potential at distances
beyond one Debye length is relatively stronger than the repulsion for larger
particles.  We exclude simulations showing extensive aggregation from further
analysis.

For small Debye lengths around \SI{10}{\nm}, the sedimentation velocity is
predicted well by Eq.~\eqref{eq:basic_batchelor} with $K\approx 6.55$.
As $\lambda_\mathrm{D}$ increases, the slope increases rapidly,
meaning that mutual hindrance is increased. While
particles close to each other sediment faster than a single particle, as shown
in Fig.~\ref{fig:particle_benchmarks}(a), at larger interparticle distances the
effect of fluid backflow dominates and particles mainly slow down each other. An
increase in $\lambda_\mathrm{D}$ leads directly to an increase in the mean distance
between next neighbors due to a longer range of the repulsive potential.
\begin{figure*}
\centering
 \includegraphics[scale=0.6]{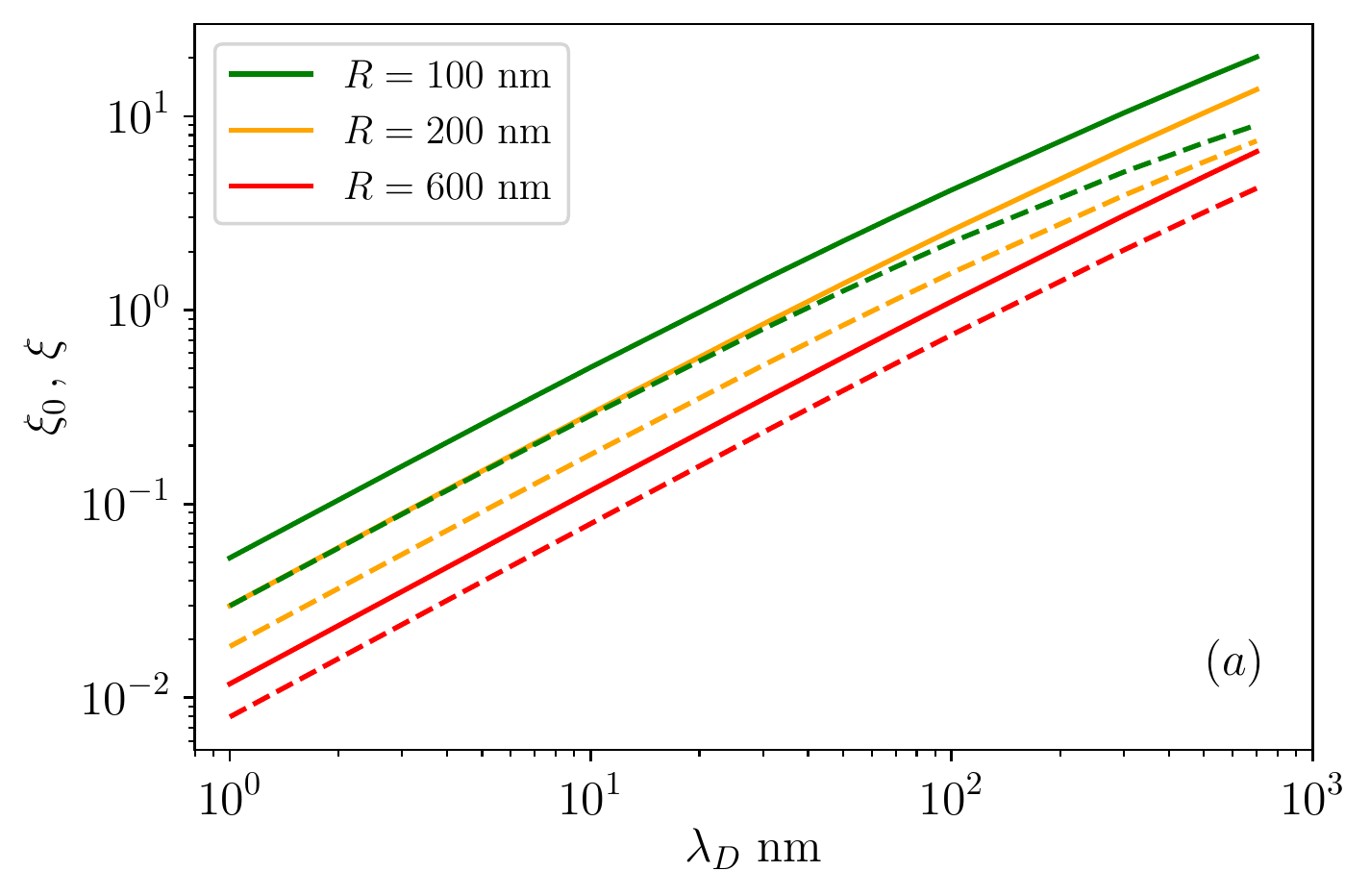} \includegraphics[scale=0.6]{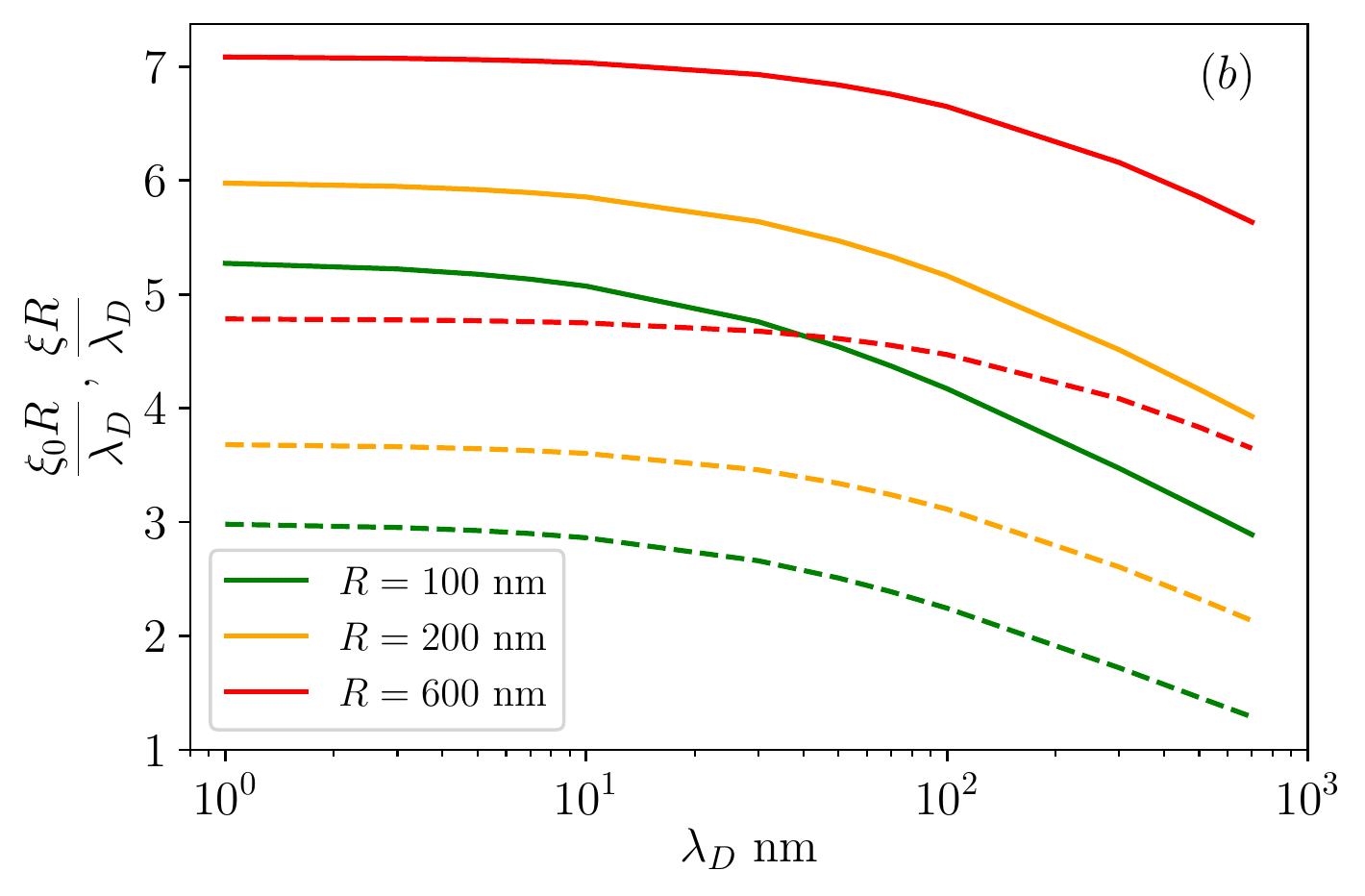}
 \caption{(a): \pacorr{Surface-surface dimensionless distance at which the
colloid-colloid DLVO interaction is comparable to the thermal energy, $\xi_0$
(defined in Eq.\eqref{eq:xi_0}, solid lines), and ten times the thermal energy,
$\xi$ (defined in Eq.\eqref{eq:xi}, dashed lines)}  as a function of the Debye
length, $\lambda_D$, for various values of the particle radius,
$R=100,200,600$nm (see legend) and for $\zeta=50$mV. (b): same data as in panel
a) but for $\xi R/\lambda_D$ and $\xi_0 R/\lambda_D$.}
\label{fig:xi} 
\end{figure*}\\

When $\lambda_\mathrm{D}$ is sufficiently large so that the particles cannot fully
escape the repulsive potential of their neighbors, $v(\phi)$ becomes distinctly
nonlinear.  The nonlinear regime begins to show in Fig.~\ref{fig:vel_phi} for
 $\lambda_\mathrm{D}=\SI{950}{\nm}$.
According to calculations by Thies et
al.,\cite{thies-weesie_nonanalytical_1995} the sedimentation behavior in the
limit of large $\lambda_\mathrm{D}$ should follow
\begin{equation}
  \label{eq:ordered_sedimentation}
\frac{v(\phi)}{v_0}=1-\varsigma\phi^{\frac{1}{\omega}},
\end{equation}
with $\varsigma \approx 1.8$ and $\omega \approx 3$. This functional form is
quite general for ordered particle
arrays.\cite{thies-weesie_nonanalytical_1995,philipse_colloidal_1997,watzlawek_sedimentation_1999}
The slope obtained from a linear fit over the concentration range
$[\phi_1,\phi_2]$ in a system described by Eq.~\eqref{eq:ordered_sedimentation}
can be predicted as
\begin{equation}
  \label{eq:ordered_slope}
  K_\omega = \varsigma \frac{\phi_2^{\frac{1}{\omega}}-\phi_1^{\frac{1}{\omega}}}{\phi_2-\phi_1}.
\end{equation}
While the more general Eq.~\eqref{eq:ordered_sedimentation} can give a better fit in
the limit of large $\lambda_\mathrm{D}$, a simpler linear fit may
often be preferable, particularly when working with data exhibiting significant
statistical errors and when varying $\phi$ across a relatively narrow range.

When $\lambda_\mathrm{D}$ is small enough for $v(\phi)$ to remain in the linear
regime, $K$ can be approximated for general interaction potentials $\Phi$ following Batchelor and Wen\cite{batchelor_sedimentation_1982} as $K=6.55-0.44\alpha$ with
\begin{equation}
  \label{eq:batchelor_potential_correction}
  \alpha=3\int_0^\infty \big( e^{-\frac{\Phi(\hat{s})}{\mathrm{k_BT}}}-1 \big) \hat{s}^2 \, \mathrm{d}\hat{s}.
\end{equation}
If the potential $\Phi$ in Eq.~\eqref{eq:batchelor_potential_correction} is set equal to
$E_\mathrm{coul}+E_\mathrm{vdw}+E_\mathrm{hs}$, the resulting $K$ depends on $R$, $\zeta$, $\lambda_D$, and $k$ in a non-trivial way. Using a more crude approximation of $\Phi$ as a step potential
$\Phi=E_0\Theta(\xi-\hat{s})$ that falls abruptly from $E_0\gg \mathrm{k_BT}$ to 0 at
surface-to-surface distance $\xi R$ one obtains the much simpler
solution\cite{batchelor_sedimentation_1982}
\begin{equation}
  \label{eq:batchelor_excluded_volume}
  K_\Phi(\xi) = 6.55 + 2.65(\xi^2+2\xi).
\end{equation}
One might be tempted to identify $\xi=\frac{\lambda_\mathrm{D}}{R}$. However, as both the radius $R$ and the zeta potential $\zeta$ influence the strength of the repulsive potential at a given distance $s / R$ significantly (see
Fig.~\ref{fig:si_reprange} in the ESI$^\dag$), we instead define $\xi$ as the surface-to-surface
distance in multiples of the radius at which $E_\mathrm{DLVO}$ first exceeds 
$\mathrm{k_BT}$ coming from infinity, thus constituting a significant potential barrier as compared to thermal energy.
The exact choice of the threshold value makes relatively little difference in
the resulting value of $\xi$ due to the fast exponential decay of the repulsive
potential.  Because this measure depends on the charge state of the particle as
well as its size, it encodes more information than the Debye length alone.
{In the dilute regime that we are focused on here, colloids are typically far apart. Hence, we can disregard the van der Waals contribution, Eq.~\eqref{eq:vdw_potential}, and we can approximate the DLVO potential by the solely electrostatic interaction, Eq.~\eqref{eq:coul_potential}. Accordingly, the surface-surface dimensionless distance,  {$\xi=\xi_0$}, at which the colloid-colloid DLVO interaction is comparable to the thermal energy is obtained by numerically solving 
\begin{align}
 4\pi R\varepsilon \zeta^2 \mathrm{e}^{-\frac{R}{\lambda_\mathrm{D}}\xi_0}=\mathrm{k_BT}(\xi_0+2)\,.
 \label{eq:xi_0}
\end{align}
Fig.\ref{fig:xi}a  shows the dependence of $\xi_0$ on $\lambda_D$ for diverse particle radii. In particular, Fig.\ref{fig:xi}a shows that $\xi_0$ can be comparable or even larger than the particle size.}

Eq.~\eqref{eq:batchelor_excluded_volume} models the effect of
the repulsive potential as an excluded volume around otherwise non-interacting
and thus randomly distributed particles. A similar approach of modelling
short-ranged DLVO interactions as an excluded volume, or alternatively an
effective particle concentration, has been used previously for example by
Gilleland et al.\cite{gilleland_new_2011} or Antonopoulou et al.\cite{antonopoulou_numerical_2018}

The impact of the repulsive potential barrier at $\hat{s} = \xi_0$ on the final particle distribution
of course depends on the average particle-particle spacing, which in turn depends on the
particle concentration $\phi$.
In order to account for this we furthermore introduce the naively calculated
average interparticle spacing
\begin{equation}
  \label{eq:interparticle_spacing}
\hat{s}_\phi = \frac{1}{R}\sqrt[3]{\frac{V_\mathrm{p}}{\phi}}-2 = \sqrt[3]{\frac{4\pi}{3\phi}}-2
\end{equation}
using the particle volume $V_\mathrm{p}$.  It is formulated in multiples of the
radius and measured from surface to surface, just like $\xi$.  Normalizing
$\xi_0$ as 
\begin{align}
\chi_0=\frac{\xi_0}{\hat{s}_\phi}
\label{eq:def_chi_0}
\end{align}
we obtain a useful dimensionless measure for the
range of the repulsive DLVO force relative to the average interparticle
distance.
{Interestingly, Fig.\ref{fig:xi}b shows that, for the values of the parameters under scrutiny, $\xi_0$ attains quite large values as compared to both $\lambda_D/R$. This implies that the relevant distance at which the colloids experience the mutual DLVO interaction can be quire larger than the Debye length hence supporting our definition of $\chi_0$ in Eq.~\ref{eq:def_chi_0}. }
\begin{figure*}[h]
  \centering
  \includegraphics{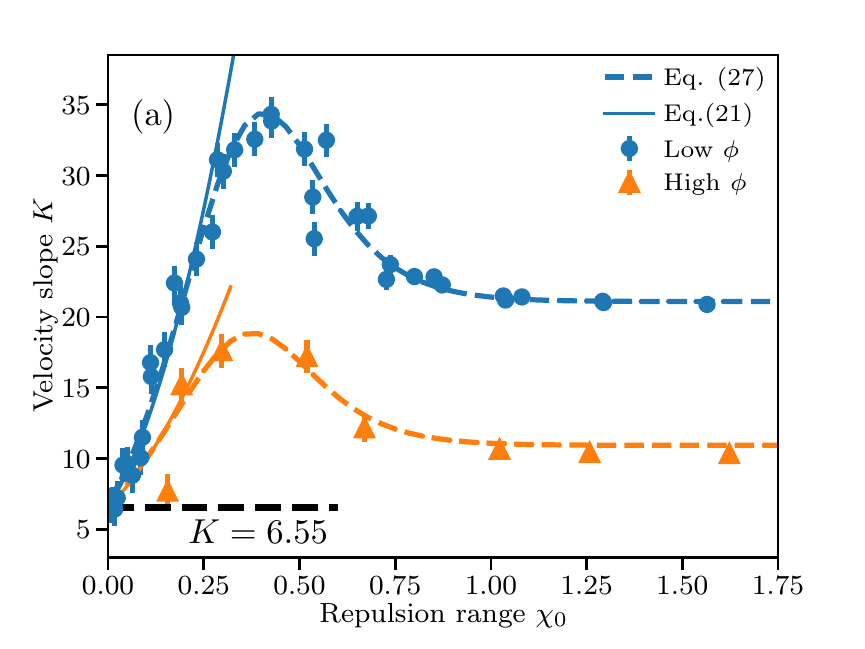}
  \includegraphics{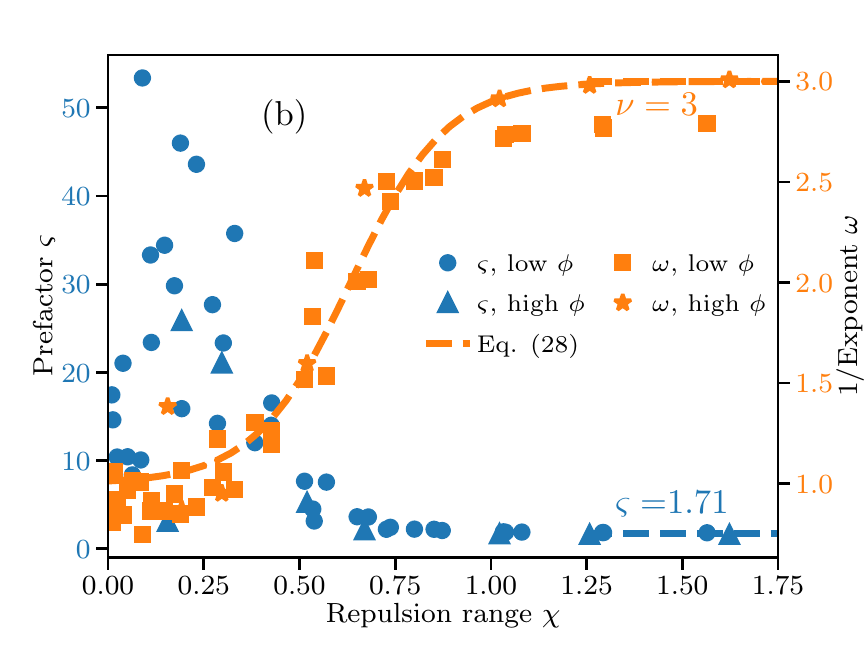}
  \caption{Linear and nonlinear fit parameters obtained for low $\phi$ ($\phi \in 0.2\%-0.8\%$) and high $\phi$ ($\phi \in 1\%-1.4\%$).
    $\chi = \xi / \hat{s}_\phi$ gives the range of the repulsive DLVO potential
    relative to the average interparticle distance.
    (a) Slope $K$ from linear fits to $v(\phi)/v_0$.
        Dashed lines follow Eq.~\eqref{eq:linslopes_fit} using identical fit parameters, and full
        lines follow Eq.~\eqref{eq:batchelor_excluded_volume}.
    (b) Parameters $\varsigma$ and $\omega$ from nonlinear fits of Eq.~\eqref{eq:ordered_sedimentation} to $v(\phi)/v_0$.
        Dashed line follows Eq.~\eqref{eq:sigmaslopes_fit}, fitted using low and high $\phi$ data combined.
        Uncertainties in the velocity $v$ translate to a large variation in $\varsigma$ for low $\chi$.
}
  \label{fig:slopes}
\end{figure*}

{We have used the surface-surface distance $\xi_0$ (see Eq.~\eqref{eq:xi_0}), and the associated value of $\chi_0$ (see Eq.~\eqref{eq:def_chi_0}), as the effective particle size in the hard-sphere model, Eq.~\eqref{eq:batchelor_excluded_volume}. However, 
the agreement is qualitatively good yet we admit some quantitative discrepancies. To address the role of the softness of the DLVO potential at distance $\xi_0$, as compared to the hard-sphere interaction, we define $\xi$ (and the associated $\chi$) as the distance at which the DLVO potential $\simeq 10 k_B T$
 by numerically solving 
\begin{align}
 4\pi R\varepsilon \zeta^2 \mathrm{e}^{-\frac{R}{\lambda_\mathrm{D}}\xi}=&10\mathrm{k_BT}(\xi+2)\label{eq:xi}\\
 \chi=&\frac{\xi}{\hat{s}_\phi}\,.
\end{align}
Indeed, at such distance the DLVO potential is stiff and therefore it may resemble the hard-sphere interaction. 
Interestingly,} the slope $K$ as a function of $\chi$ in Fig.~\ref{fig:slopes}(a) neatly
collapses onto a single curve when calculated over a fixed range of
concentrations {hence showing that $\chi$ is the dimensionless variable that captures the ``dynamics''.}


The error bars in Fig.~\ref{fig:slopes}(a) account for variations due to the randomness involved
in initial particle placement and the subsequent equilibration of particle distributions under thermal fluctuations.
To estimate the error bars we repeat simulations up to 6 times at selected parameter combinations spanning the whole range of $\chi$ with different
random number seeds and calculate the standard deviation of
the resulting velocities as described in section~\ref{sec:si_errors} of the ESI$^\dag$.
The error bars strongly depend on $\chi$ and are largest for
non-interacting particles.
Knowing $\chi$ and the concentration
at which $K$ is measured, we can predict the value of $K$ to a decent accuracy
both for small (via Eq.~\eqref{eq:batchelor_excluded_volume}) and large $\chi$ (via Eq.~\eqref{eq:ordered_slope}).
At intermediate $\chi$ an interpolating fit drawn in dashed lines in Fig.~\ref{fig:slopes}(a) matches the observed trend well.\\
This {interpolating} fit {captures the} transition from $K_\Phi(\xi)$ (Eq.~\eqref{eq:batchelor_excluded_volume}) to the constant value $K_\omega$ given by Eq.~\eqref{eq:ordered_slope} via a sigmoid function
\begin{equation}
  \label{eq:linslopes_fit}
  {K(\chi) = 2\dfrac{K_\Phi(\xi) - K_\omega}{1 + \exp\left(\frac{\chi - \chi_\mathrm{m}}{\delta_K} \right)} + K_\omega.}
\end{equation}
{We remark that for $\chi\rightarrow \chi_m$ Eq.~\eqref{eq:linslopes_fit} gives $K\simeq K_\Phi(\xi)$ whereas for $\chi\rightarrow \infty$ we get $K\simeq K_\omega$.}
We determine the two fit parameters $\chi_\mathrm{m} \approx 0.38$ - roughly corresponding to the position of the maximum, and $\delta_K \approx 0.096$ - giving the scale in $\chi$ over which the transition to a locally ordered suspension occurs, from a combined fit to the complete data set for both $\phi = 0.2-0.8\%$ and $\phi = 1-1.4\%$.
The dashed lines in Fig.~\ref{fig:slopes}(a) both follow Eq.~\eqref{eq:linslopes_fit} using the same values of $\chi_\mathrm{m}$ and $\delta_K$.
The two lines differ only due to the different values of $\phi_{^1\!/\!_2}$ used in calculating $K_\omega$ and the different values of $\xi = \chi \hat{s}_\phi$ inserted in $K_\Phi$ for a given $\chi$, again due to the different volume fractions $\phi$.
In calculating $K_\omega$ we set $\omega = 3$ and $\varsigma \approx 1.71$ regardless of $\phi$.
\pacorr{We remark that the value} $\varsigma \approx 1.71$ \pacorr{has been} obtained by averaging over all data points from nonlinear fits to Eq.~\eqref{eq:ordered_sedimentation} at $\chi > 1.2$, as indicated in Fig.~\ref{fig:slopes}(b).

One can reformulate the fitted $K(\chi)$ from Eq.~\eqref{eq:linslopes_fit} as a function of $\phi$ for fixed $\xi$ and perform numerical integration to reconstruct the hindrance function $v(\phi)/v_0$ as shown in section~\ref{sec:si_velintegration} in the ESI$^{\dag}$.\\
As shown in Fig.~\ref{fig:slopes}(a) we recover the case of non-interacting particles for $\chi \to 0$ and $K\to
6.55$ as in Eq.~\eqref{eq:basic_batchelor}.
Up to $\chi\approx 0.3$, $K$ is well-approximated by
Eq.~\eqref{eq:batchelor_excluded_volume}, which is shown as full lines in
Fig.~\ref{fig:slopes}(a). Eq.~\eqref{eq:batchelor_excluded_volume} fails as a
valid approximation when the particle distribution cannot be approximated as
homogenous in space, i.e. when the RDF deviates from the step function expected
for dilute hard spheres with an effective radius increased by $R\xi/2$.

Fig.~\ref{fig:slopes}(b) shows the obtained parameters $\varsigma$ and $\omega$ from nonlinear fits to Eq.~\eqref{eq:ordered_sedimentation}.
While the nonlinear fit works well in the locally ordered regime at
$\chi\gtrsim 0.4$ and $\omega \approx 1$ is correctly reproduced even for $\chi
\to 0$, there is a large uncertainty in $\varsigma$ at $\chi\lesssim 0.4$. As
shown in Fig.~\ref{fig:particle_benchmarks}(b) and (c), the uncertainty in the
velocity in the disordered regime is much larger than in the locally ordered regime.
According to Eq.~\eqref{eq:ordered_sedimentation}, $v/v_0$ depends much more sensitively on the exponent $\omega$ than on the prefactor $\varsigma$, in particular when $\phi$ is small.
This can be seen from the ratio of the derivatives $(\partial v / \partial \varsigma) / (\partial v / \partial \omega) = - \omega^2 /{(} \varsigma \ln \phi{)}$, which goes to zero for small $\phi$.
Accordingly, uncertainties in the velocity translate into much larger uncertainties in the values of $\varsigma$ than of $\omega$, leading to the large spread in the obtained values of $\varsigma$ at low $\chi$ in Fig.~\ref{fig:slopes}(b).
The exponent $\omega$ can be predicted well from $\chi$ via a fitted sigmoid function
\begin{equation}
  \label{eq:sigmaslopes_fit}
  {\omega(\chi) = 2 \dfrac{1}{1+\exp\left(-\frac{\chi - \chi_0}{\delta_\omega}\right)} + 1,}
\end{equation}
with the fit parameters $\chi_0 \approx 0.63$ and $\delta_\omega \approx 0.13$ again obtained via a single fit to the combined data for all $\phi$ as in Eq.~\eqref{eq:linslopes_fit}.

\begin{figure*}
  \centering
  \includegraphics{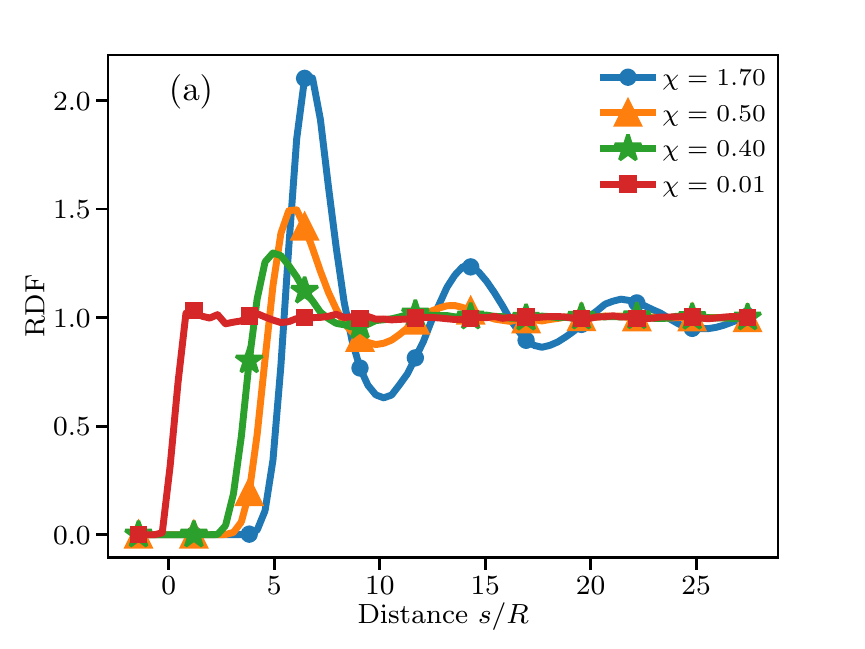}
  \includegraphics{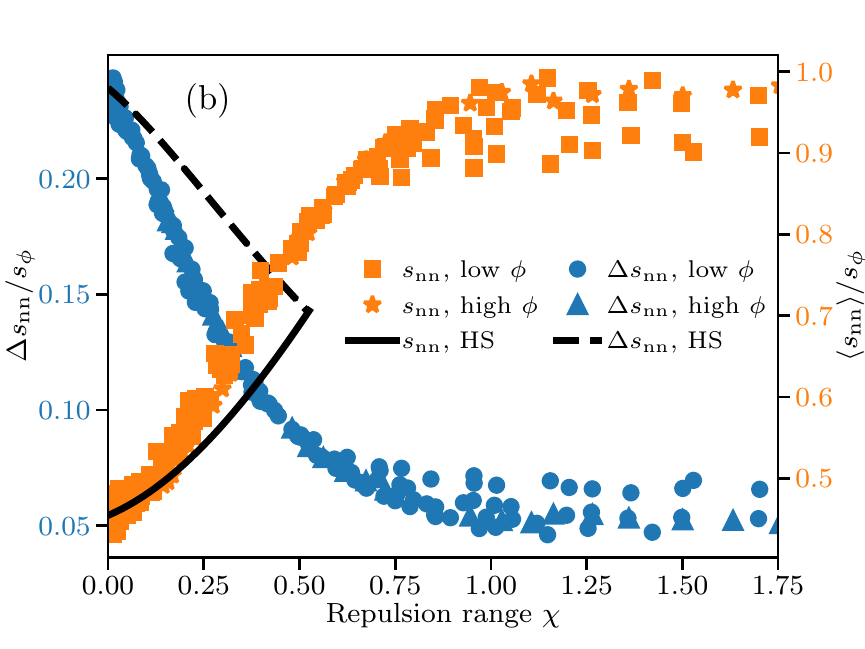}
  \caption{
    (a) Changes in radial distribution function induced by repulsive DLVO
        interactions at $\phi=0.8\%$.
    (b) Distance to next-neighbor-particle $s_{\mathrm{nn}}$ and its standard
        deviation $\Delta s_{\mathrm{nn}} = \sqrt{\mathrm{Var}(s_{\mathrm{nn}})}$ normalized by $s_\phi$.
        Black lines give theoretical results for homogenous particle distributions with an infinite step potential effectively extending the hard sphere radius in proportion to $\chi$ by $\xi R / 2$.}
  \label{fig:particle_dist}
\end{figure*}

Fig.~\ref{fig:particle_dist}(a) shows the changes in the RDF leading in turn to the changes in $K$ and $\omega$.
For small $\chi$ the RDF is a simple step function as expected for randomly distributed hard spheres.
The first deviation from this idealized form is visible for $\chi=0.4$ in the form of a pronounced primary maximum next to the exclusion zone.
The transition from a disordered to a locally ordered particle distribution is
accompanied by an oscillatory component in the RDF, which becomes visible at
$\chi \approx 0.5$.
The length scale over which the oscillations decay can be interpreted as the length scale over which particle positions are correlated.
Predictably, this length scale increases markedly as $\chi$ increases, with the RDF for $\chi=1.7$ showing visible correlation at distances well beyond 25 particle radii.
Our results here qualitatively agree well with the RDF of charged sphere suspensions obtained in other works.\cite{gilleland_new_2011,banchio_short-time_2008,gapinski_collective_2007,thies-weesie_nonanalytical_1995}

Going to higher values of $\phi$ or $\chi$ than those we studied should lead to crystalline bcc or fcc particle distributions.\cite{tata_ordering_2006,monovoukas_experimental_1989}
Simulating crystalline systems would require great care though, because their very long-ranged order may be strongly affected by finite system sizes and take a long time to equilibrate.\cite{HHB08}
Furthermore, DLVO force models may be ill-suited for such systems, as they fail to properly model the experimentally observed coexistence of colloidal crystals with disordered phases in dilute suspensions at small salt concentration.\cite{ise_structure_2005}

In Fig.~\ref{fig:particle_dist}(b) we can see how changes in $\chi$ affect the average surface-to-surface interparticle distance $\langle s_{\mathrm{nn}} \rangle$ as well as its standard deviation $\Delta s_{\mathrm{nn}}$ within the particle configuration used in a given hydrodynamic simulation.
Due to normalization by the average interparticle spacing $s_\phi$ the results for different $\phi$ again collapse rather well on a single curve.
Unlike in Figs.~\ref{fig:slopes}(a) and (b), where each data point represents a group of simulations at different $\phi$ and otherwise identical parameters (with $\chi$ being averaged over $\phi$), each data point here corresponds to a single simulation.

The full and dashed lines in Fig.~\ref{fig:particle_dist}(b) compare the simulation results with a homogenous suspension of particles interacting only via the step potential $\Phi = E_0 \Theta(\xi - \hat{s})$ with $E_0 \to \infty$ at $\hat{s} = \xi$ - like hard spheres with a radius enlarged in proportion to $\chi$.
Because the particle distribution (derived in section~\ref{sec:si_neighbors} of the ESI$^\dag$) neglects particle-particle correlations beyond the range of the step potential, it is neccessarily inaccurate when either $\phi$ or $\chi$ are large.

A substantial discrepancy between the simulation results for $\langle s_{\mathrm{nn}} \rangle$ and the hard sphere distribution develops starting around $\chi = 0.3$.
This is consistent with the observation that the enlarged hard sphere model from Eq.~\eqref{eq:batchelor_excluded_volume} predicts $K(\chi)$ well only up to $\chi \approx 0.3$, as shown in Fig.~\ref{fig:slopes}(a).
Interestingly, $\Delta s_{\mathrm{nn}}$ seems to diverge from the hard sphere distribution much faster though, showing that the microstructure of the DLVO suspension does differ noticeably from the hard sphere suspension for $\chi < 0.3$, despite affecting sedimentation in much the same way as an increased hard sphere radius.

For $\chi\gtrsim 1$, $\langle s_{\mathrm{nn}} \rangle$ approaches the maximal average interparticle distance $s_{\phi}$ and $\Delta s_{\mathrm{nn}}$ indicates a narrow distribution of next-neighbor distances as expected for a locally ordered particle distribution.

\section{Conclusions}
By simulating the hydrodynamic and DLVO interactions of large ensembles of
particles we found seemingly universal trends in the sedimentation behavior for
a wide range of Debye lengths and particle sizes. We quantified the effect of
particle interactions depending on the range $\chi$ of electrostatic repulsion
in our results either via the slope $K$ or the exponent $1 / \omega$ of fits to
the sedimentation velocity $v(\phi)/v_0$ across different ranges of $\phi$.

$K$ is the slope extracted from a linear fit to $v(\phi)/v_0$ and appears to be
described well by our fit to Eq.~\eqref{eq:linslopes_fit} for any $\phi$ in the
dilute limit.  Eq.~\eqref{eq:linslopes_fit} predicts $K(\chi)$ assuming that
the electrostatic repulsion at $\chi \lesssim 0.3$ acts merely like an increase
of the effective hard sphere radius, whereas at $\chi \gtrsim 1$ sedimentation
follows the known solution Eq.~\eqref{eq:ordered_sedimentation} with $\varsigma
\approx 1.71$ and $\omega=3$ for ordered particle arrays.  The transition from
one solution to another is approximated in Eq.~\eqref{eq:linslopes_fit} using a
simple sigmoid function.

Applying non-linear fits following Eq.~\eqref{eq:ordered_sedimentation} to
$v(\phi)/v_0$ instead we find clear nonlinearity ($\omega > 1$) commencing
around $\chi = 0.4$, where $K(\chi)$ reaches its maximum.  Near this transition
point from linear to nonlinear the RDF shows a transition from a disordered
gas-like state to a liquid-like state.  $\omega(\chi)$ is likewise describable
by a sigmoid function, with a smoothened step-like transition from $\omega = 1$
at $\chi \to 0$ to $\omega = 3$ at $\chi \gtrsim 1$. This coincides with the
point where the average next-neighbor distance reaches its maximum possible
value.

Both $K(\chi)$ following Eq.~\eqref{eq:linslopes_fit} and $\omega(\chi)$
following Eq.~\eqref{eq:sigmaslopes_fit} offer themselves as a potentially
useful gauge to estimate the extent of electrostatic interactions (encoded by
$\chi$) in a suspension directly from experimental measurements of the
sedimentation velocity under varied particle concentration.  The approach of
estimating $\chi$ via $\omega$ has the advantage that $\omega$ increases
monotonously with $\chi$ and hence can in principle be inverted to obtain a
mapping $\chi(\omega)$. The downside of this approach is that the nonlinear
fits tend to be more sensitive to noise in $v(\phi)$ than linear fits.

We note in conclusion that our results, while obtained under the assumption of
a strongly repulsive DLVO potential at $\zeta=\SI{50}{\milli\V}$, are in fact
generally valid for any repulsive potential with a steep potential barrier at
distance $\xi=\chi\hat{s}_\phi$. The van der Waals interactions are strongly
subdued in most of our parameter regime and the models we used to predict $K$
at both small and large $\chi$ are not specific to details of the DLVO
interaction.

In future work we aim to reproduce long-ranged electrostatic interactions in
sedimentation velocity experiments for a broad parameter range and compare the
experimental data directly to our simulations. In the experimental setup we
wish to study model nanoparticle systems including a controlled degree of
polydispersity. Other possible avenues of future research might include
non-spherical, in particular rod-like charged particles, where orientation and
rotation become important in addition to translational ordering.

\section*{Conflicts of interest}
There are no conflicts to declare.

\section*{Acknowledgements}
This work \pacorr{has been} supported by the Competence Network for Scientific and Technical High Performance Computing in Bavaria (KONWIHR) and \pacorr{has been} further funded by the Deutsche Forschungsgemeinschaft (DFG, German Research Foundation) -- Project-ID 416229255 -- SFB 1411.






\providecommand*{\mcitethebibliography}{\thebibliography}
\csname @ifundefined\endcsname{endmcitethebibliography}
{\let\endmcitethebibliography\endthebibliography}{}


\end{document}